\documentclass[prd,letterpaper,preprint,showpacs,superscriptaddress,floatfix,nofootinbib]{revtex4}

\usepackage{graphicx,psfrag,amsmath,amssymb,amsfonts,bbm,latexsym,color,dcolumn,bm}

\begin{document}

\title{High-energy density implications of a gravitoweak unification scenario}

\author{Roberto Onofrio}
\email{onofrior@gmail.com}
 
\affiliation{Dipartimento di Fisica e Astronomia 'Galileo Galilei', 
Universit\`a di Padova, Via Marzolo 8, Padova 35131, Italy}

\affiliation{ITAMP, Harvard-Smithsonian Center for Astrophysics, 
60 Garden Street, Cambridge, MA 02138, USA}
\date{\today}

\begin{abstract}
We discuss how a scenario recently proposed for the morphing of 
macroscopic gravitation into weak interactions at the attometer 
scale affects our current understanding of high-energy density phenomena. 
We find that the Yukawa couplings of the fundamental fermions are directly related to 
their event horizons, setting an upper bound $y_f\leq \sqrt{2}$ for their observability 
through gauge interactions. Particles with larger Yukawa couplings are not precluded, but 
should interact only gravitationally, providing a natural candidate for dark matter. 
Furthermore, the quantum vacuum contribution to the cosmological constant is 
reduced by several orders of magnitude with respect to the current estimates. 
The expected running of the Newtonian gravitational constant could provide 
a viable alternative scenario to the inflationary stage of the Universe.
\end{abstract}


\maketitle
Recently, we have proposed a conjecture in which weak interactions should be considered 
as the microscopic counterpart of macroscopic gravitation \cite{Onofrio}.  
This conjecture relies upon a quantitative relationship between the Fermi constant of 
weak interactions $G_F$ and a renormalized Newtonian universal gravitational constant 
$\tilde{G}_N$ which satisfy\footnote{This relationship differs by a factor 2 
from Eq. 2 in \cite{Onofrio} since the Planck mass has been chosen as the one 
corresponding to the equality between the Compton wavelength and the Schwarzschild 
radius, {\it i.e.} $\hbar/(M_P c)=2 G_N M_P/c^2$, the factor 2 in the Schwarzschild 
radius having being instead omitted in \cite{Onofrio}.}
\begin{equation}
G_F=\sqrt{2}\left(\frac{\hbar}{c}\right)^2 \tilde{G}_N.
\end{equation}
This expression holds provided that we choose $\tilde{G}_N=1.229
\times 10^{33} G_N=8.205 \times 10^{22}~\mathrm{m^3~ kg^{-1}~ s^{-2}}$.
Models for a gravitoweak unification have been discussed in the past 
using formal arguments \cite{Hehl,Hehl1,Batakis,Loskutov,Capozziello,Alexander}, 
and a possible running of the Newtonian gravitational constant in purely
four-dimensional models has been recently proposed \cite{Capozziello,Calmet}.
In \cite{Onofrio}, we have focused the attention on consequences
of this conjecture at low and intermediate energies, with particular
emphasis on anomalous gravitational contributions to bound states with
size in the femtometer to picometer distance range. 
In following contriibutions, we have developed a quantitative attempts to explain 
the so-called ``proton radius puzzle'' arising from high-precision spectroscopy 
on muonic hydrogen \cite{Muonich}. We discuss here features of this morphing 
with implications for scattering states such as in high-energy experiments at 
colliders and in cosmic rays observatories, and finally comment on its interplay 
with the cosmological constant problem. This contribution should be considered as 
part of a more extended program, pioneered by Amelino-Camelia and collaborators, 
whose aim is to develop quantum gravity with a strong phenomenological content, 
allowing for a faster constructive feedback from Nature on the otherwise virtually 
infinite ways to unfold this subfield if no connection to observable reality is actively 
pursued \cite{Amelino}.

The starting point of our analysis is that if the relevant parameter describing 
gravity in the microworld is the renormalized Newtonian gravitational constant 
$\tilde{G}_N$, then the Schwarzschild radii of the elementary particles 
become quantities of phenomenological interest, as they are boosted 
by 33 orders of magnitude with respect to the standard scenario using
$G_N$. For instance, the Schwarzschild radius of the top quark should
become ${R_s}^{(t)}=2 \tilde{G}_N m_t/c^2=5.67 \times 10^{-19}$ m, a lengthscale 
not far from those currently explored at the highest energy colliders. 
Insights on the physical consequences of these large Schwarzschild radii 
may be gained by considering a dimensionless parameter 
$\eta_f= {R_s}^{(f)}/\tilde{\Lambda}_P$, allowing to compare the 
Schwarzschild radius of a generic fundamental fermion 
$R_s^{(f)}=2 \tilde{G}_N m_f/c^2$ to the lengthscale of quantum vacuum 
fluctuations of space-time, {\it i.e.} the renormalized Planck length 
$\tilde{\Lambda}_P=(2\hbar \tilde{G}_N/c^3)^{1/2}=8.014 \times 10^{-19}$ m. 
Considering that the renormalized Planck energy $\tilde{E}_P$ coincides with 
the Higgs vacuum expectation value $v$ as commented in \cite{Onofrio}, 
$\tilde{E}_P=(\hbar c^5/(2 \tilde{G}_N))^{1/2}=v$, we have the following chain of 
relationships

\begin{equation}
\eta_f=\frac{R_s^{(f)}}{\tilde{\Lambda}_\mathrm{P}}=
\left(\frac{2\tilde{G}_N}{\hbar c}\right)^{1/2}\hspace*{-0.4cm}m_f=
\left(\frac{{\sqrt{2}c G}_F}{\hbar^3}\right)^{1/2}\hspace*{-0.4cm}m_f=
\frac{m_f c^2}{v}=\frac{y_f}{\sqrt{2}},
\end{equation}
where we have introduced the Yukawa coupling coefficient $y_f$ such that 
the mass of a fundamental fermion is written as $m_f=y_f v/(\sqrt{2}c^2)$.
We assume that the parameter $\eta_f$ determines the possibility for a fundamental 
fermion to communicate information via gauge bosons only if its horizon is smaller than 
the renormalized Planck length, that is $\eta_f \leq 1$. 
In this framework it becomes natural that the Yukawa coefficients are 
small, as  $R_s^{(f)}<\tilde{\Lambda}_\mathrm{P}$ implies $y_f<\sqrt{2}$ for 
all particles observable through their non-gravitational effects. 
If the Schwarzschild radius exceeds the Planck length, {\it i.e.} $y_f >\sqrt{2}$, 
a particle should behave like a black hole, therefore gauge bosons emitted and reabsorbed 
via virtual vacuum effects will be unable to propagate beyond the 
horizon, and only gravitational effects may be detectable, such as the ones 
currently attributable to dark matter.\footnote{If the Compton wavelength of a gauge boson 
is much larger than the Schwarzschild radius of the particle emitting or absorbing 
the gauge boson, the latter does not play any role for the boson-exchange, as in 
usual quantum field theory in flat spacetime, and the range of the mediated interaction 
is established by the Compton wavelength, for instance infinite raneg in the case of 
the photon, attometer range in the case of the $W^{\pm}$ and $Z^0$ bosons, so gauge bosons 
do not have any horizon limitation for their virtual propagation. In the opposite situation, 
instead, we expect the emission of virtual gauge bosons to be strongly suppressed and 
limited to a range determined by the Schwarzschild radius of the involved particle, rather 
than the Compton wavelength of the boson itself, as well-known for the simplest, celebrated 
case of photons which allowed to define black holes themselves. The possibility of emitting gauge 
bosons via Hawking radiation is not considered here since for particles with mass 
$m > v/c^2$ the temperature of the Hawking radiation is not consistently defined 
(or, equivalently, the Hawking radiation may be reliably estimated only for energy 
scales much smaller than the one of quantum gravity, which in our setting means 
for Hawking temperatures $T_H<<v/K_B$). On the stability of black holes at the 
Planck scale see \cite{Markov,Bunch,Mukhanov,MacGibbon,Frolov}.} 

This has direct implications for high-energy physics experiments since the discovery 
of new massive states of mass larger than $246.22$ GeV/$\mathrm{c^2}$, if using 
high-energy physics detectors relying on non-gravitational interactions, should be precluded. 
Moreover, even existing fundamental fermions will be affected by this constraint at energies large
enough. The minimum distance between two colliding particles is related to the 
center-of-mass energy $E_{cm}$, in the relativistic limit, as $\delta x \simeq h c/E_{cm}$, {\it i.e.} the 
corresponding de Broglie wavelength. If this distance is smaller than the Schwarzschild radius 
the particles will be confined within the horizon and afterward will not be able to propagate 
beyond it apart from subleading Hawking effect propagation via quantum tunneling. 
This may happen in spite of the fact that the energy-momentum dispersion relationship 
is violated, something absolutely possible in a quantum gravity regime in which we do not expect 
necessarily on-shell states. Therefore, for collisions between fundamental particles with 
center-of-mass energy larger than $E_s^{(f)}=h c/R_s^{(f)}$ we expect inhibition of scattering at large 
momentum transfers of order $q^2 \simeq (\hbar/R_s)^2$. 

As seen in Table I, this progressively affects particles of decreasing mass, so
the first to be hit by this inhibition should be the top quark. 
This could be seen, in the cleanest environment provided by an
electron-positron collider (or an equivalent $\mu^+ \mu^-$ machine) 
in the 2.2 TeV range, as a sudden drop in the normalized cross-section 
$\sigma(e^+e^-\rightarrow \mathrm{hadrons})/\sigma(e^+e^-\rightarrow \mu^+\mu^-)$ 
from its expected value (without QCD corrections) of 5 in the 
range 175 GeV $<E_\mathrm{cm}<$ 2.2 TeV, to a value of 11/3 for
$E_\mathrm{cm}>$ 2.2 TeV, since the top quark is no longer an active degree of freedom. 
Other possible implications are in the high energy component of the
cosmic rays and in high-energy neutrinos observatories, with limitations 
to very high-energy flux. For instance, muons with energy $E_{thr}^{\mu}=
{E_s}^2/(2m_p c^2) \simeq 7 \times 10^{12}$ GeV will reach their event horizon 
limit when interacting with protons, and therefore we expect inhibition of their measurable 
flux for energies higher than the threshold energy $E_{thr}^{\mu}$. 
We also notice, see last column in Table I, that the Schwarzschild energy density of 
fermions is inversely proportional to the square of their mass, so lightest particles 
have higher Schwarzschild energy density than heaviest ones.
This is suggestive of interpreting the weak decays of fundamental fermions as induced by 
gravitational collapses to states with progressively larger gravitational binding energy. 

\begin{table}[t]
\begin{center}
\begin{tabular}{|l|c|c|c|c|c|}
\hline
 & $m_f$ & $R_s^{(f)}$ (m) & $R_s^{(f)}/\tilde{\Lambda}_\mathrm{P}=y_f/\sqrt{2}$ & $E_s^{(f)}=h c/R_s^{(f)}$ (TeV) & $\rho_s^{(f)}=m_fc^2/V_s^{(f)}$ ($\mathrm{TeV/am^3}$) \\ 
\hline
$e$     & 0.511 MeV/c${}^2 $   & $1.66 \times 10^{-24}$  & $2.07 \times 10^{-6}$ & $7.45 \times 10^5$   & $2.65 \times 10^{10}$  \\
$\mu$   & 105.66 MeV/c${}^2$   & $3.44 \times 10^{-22}$  & $4.29 \times 10^{-4}$ & $3.60 \times 10^3$             & $6.20 \times 10^{5}$   \\
$\tau$  & 1.777 GeV/c${}^2 $   & $5.78 \times 10^{-21}$  & $7.21 \times 10^{-3}$ & $214.29$              & $2.19 \times 10^{3}$   \\
\hline
$u$     & 2.75 MeV/c${}^2$     & $8.95 \times 10^{-24}$  & $1.12 \times 10^{-5}$ & $1.38 \times 10^5$    & $9.16 \times 10^{8}$   \\
$d$     & 5.5 MeV/c${}^2$      & $1.79 \times 10^{-23}$  & $2.23 \times 10^{-5}$ & $6.92 \times 10^4$   & $2.29 \times 10^{8}$   \\
$s$     & 95  MeV/c${}^2$      & $3.09 \times 10^{-22}$  & $3.86 \times 10^{-4}$ & $4.01 \times 10^3$             & $7.67 \times 10^{5}$   \\
$c$     & 1.25 GeV/c${}^2$     & $4.06 \times 10^{-21}$  & $5.08 \times 10^{-3}$ & $304.63$              & $4.43 \times 10^{3}$   \\  
$b$     & 4.70 GeV/c${}^2$     & $1.53 \times 10^{-20}$  & $0.019$              & $81.02$              & $3.14 \times 10^{2}$   \\
$t$     & 174.2 GeV/c${}^2$    & $5.67 \times 10^{-19}$  & $0.707$              & $2.186$               & $0.228$  \\
\hline
\end{tabular}
\caption{Impact on the Schwarzschild radius $R_s$ of the renormalized
Newtonian constant of gravitation $\tilde{G}_\mathrm{N}$. 
In each row we report the fundamental fermions with known mass (no neutrinos), the 
corresponding value of the Schwarzschild radius, its ratio to the renormalized Planck 
length $\tilde{\Lambda}_\mathrm{P}$, the value of the energy $E_s$ at which the Schwarzschild radius becomes 
equal to the de Broglie wavelength, and the Schwarzschild energy density of the fermion 
$\rho_s^{(f)}$, assuming that it is confined in a spherical volume $V_s^{(f)}$ of 
radius $R_s^{(f)}$, with energy in units of TeV and distance in units of attometer.
The top quark, due to its large mass and proximity to the renormalized Planck scale, 
is the most promising one to look for effects related to the event horizon in collider physics.}
\end{center}
\end{table}

The discussion above should be complemented by considering, for charged fermions, the 
Reissner-Nordstr\"om radius defined as $R_q=(q^2 \tilde{G}_N/(4\pi\epsilon_0c^4))^{1/2}$.
The condition for not emitting gauge bosons for a charged particle is obtained 
whenever the Planck length is smaller than either the Schwarzschild or the Reissner-Nordstr\"om radii. 
The Reissner-Nordstr\"om radius is equal to $R_s=q \times 4.83 \times 10^{-20}$ m where $q$ is expressed 
in elementary charge units, so $q=(0, \pm 1/3, \pm 2/3, \pm 1)e$ for fundamental fermions. 
This gives a condition on the fermion mass, and in particular, it turns out that $R_s>R_q$ 
for a mass $M>qM_0/2$, where $M_0$ is the mass for which the renormalized gravitational 
interaction equals the electromagnetic interaction for a particle endowed with 
an elementary charge unit, {\it i.e.} such that $\tilde{G}_N M_0^2=e^2/(4\pi \epsilon_0)$, 
corresponding to $M_0={[e^2/(4\pi \epsilon_0 \tilde{G}_N)]}^{1/2}=$ 31 GeV.
This means that $R_s>R_q$ for quarks with charge $q=\pm $ 1/3  if $M>$5.16 GeV/c${}^2$, 
for quarks with $q=\pm$ 2/3 if $M>$10.33 GeV/c${}^2$, and for charged leptons with charge  
$q=\pm$ 1 if $M>$15.5 GeV/c${}^2$, where $q$ expressed in elementary charge units. 
The relevant event horizon is $R_s$ for the top quark alone, while 
all other quarks and charged leptons have $R_s<R_q$, and the neutrinos should obviously have $R_q=0$.

We now sketch some considerations on the interplay between the renormalized 
Newtonian constant $\tilde{G}_N$ and the cosmological constant.
As already outlined in \cite{Onofrio} the identification of the Fermi constant with 
a renormalized Newtonian universal constant via fundamental constants $\hbar$ and 
$c$ allow to identify Fermi and Planck scales as identical, $\tilde{E}_\mathrm{P}=v$, 
where $\tilde{E}_\mathrm{P}$ and $v$ are respectively the renormalized Planck constant 
and the vacuum expectation value of the Higgs field, sometimes called the Fermi scale, 
then avoiding any hierarchy issue. Another longstanding issue of the current interface 
between cosmology and high-energy physics - the contribution of quantum fluctuations 
to the cosmological constant - is also mitigated. With the new definitions of the 
Planck energy and the Planck length $\tilde{\Lambda}_\mathrm{P}=\hbar c/v$, the vacuum
energy density at the Planck scale is written as
\begin{equation}
\rho_\mathrm{P}=\frac{\tilde{E}_\mathrm{P}}{{\tilde{\Lambda}_\mathrm{P}}^3}=
\frac{c^7}{4\hbar{\tilde{G}_N}^2}=\frac{v^4}{(\hbar c)^3}=
0.48~ \mathrm{TeV/am^3},
\end{equation}
which represents the ratio between the vacuum expectation energy of the 
Higgs field $v$ and a confinement volume with size equal to the associated 
Compton wavelength $\hbar c/v$. Due to the larger value of $\tilde{G}_N$, 
this results in a vacuum density at the Planck scale $1.5 \times 10^{66}$ smaller 
than the one estimated with the currently assumed Planck scale using the macroscopic, 
measurable value of the Newtonian constant. This is still a large contribution both with
respect to a zero value for a null cosmological constant and with respect to the value 
inferred from the interpretation of the data on the SNIa events in terms of an accelerating 
Universe (see however \cite{Sankar,Mattsson} for a critical analysis of this
interpretation). Nevertheless, the issue of the largeness of the cosmological 
constant contribution due to the quantum fields at least now coincides with the 
issue of the largeness of the vacuum density energy due to the Higgs field \cite{Weinberg,Okopinska,Shapiro}. 
It is possible, as remarked in \cite{Shapiro}, that a large nonminimal 
coupling  between the Higgs and the space-time curvature of the form 
$\xi R \phi^2$, with $\xi$ the related coupling constant, $R$ the Ricci 
scalar and $\phi$ the Higgs field, may conspire to end up with an effective 
cosmological constant much smaller then the one evaluated above (see 
\cite{OnofrioHiggs}, \cite{Onofriogravity}, and \cite{CalmetHiggs} for 
first attempts to determine bounds of curvature-Higgs couplings in astrophysics, 
gravitation, and high-energy physics, respectively). 

Finally, we compare our insights with two relevant directions already 
pursued, emphasizing analogies and differences.
First,  intriguing scenarios in which the Newtonian gravitational 
constant becomes instead zero at small distances have been proposed 
\cite{Markov1,Bonanno,Reuter1}. These models remove any issue of 
compatibility of structureless, point-like particles with a finite 
horizon since the Schwarzschild radius goes to zero along with the 
Newtonian constant \cite{Ward1,Ward2,Ward3}. 
While predictions and their observation will be required to rule 
out either this asymptotic freedom scenario or the one presented in 
this paper, we notice that if setting a null Newtonian gravitational constant 
at small distances an explanation of the tentative acceleration of the
Universe seems rather unnatural, since at earlier times the retaining 
effect of gravity would have been weaker than at later times. 
Vice versa, in our setting tentative accelerating stages of the 
Universe such as inflation could be at least partially attributed 
to a progressive decrease of the Newtonian gravitational constant 
as the expansion goes on, with the decreased gravitational retention 
mimiking an effective acceleration. Second, in a series of pioneering papers, 
Markov has discussed the possibility of an upper bound to the mass of elementary 
particles (so-called {\sl maximons}) \cite{Markov2} and its cosmological 
consequences \cite{Markov3,Markov4}. This direction has been later pursued 
by Kadyshevsky and collaborators \cite{Kadyshevsky1,Kadyshevsky2,Kadyshevsky3,Kadyshevsky4}. 
While sharing some features, in our proposal the Planck scale is shifted 
down at the Fermi scale, and we do not interpret this scale as an upper 
bound to the mass of elementary particles, rather as a bound to the 
mass of particles with other than gravitational interactions. 

In conclusion, we have performed a preliminary analysis of what we consider 
the most prominent high-energy implications of a Newtonian gravitational 
constant increasing at small distances and morphing into what we call 
weak interactions at a Planck scale coinciding with the Fermi scale. 
Among the appealing features emerging from this scenario are the understanding 
of the observed small Yukawa couplings in terms of event horizons, the absence 
of a hierarchy problem, and the related mitigation of the cosmological constant problem to the Higgs scale. 
We have also discussed predictions for the non-observability at LHC energies of particles 
with mass larger than the Higgs VEV, in particular the inhibition of gauge boson-mediated 
scattering from existing particles above well-defined thresholds corresponding to the Schwarzschild 
energy. In this scenario the top quark has a privileged status since it should be 
the first to be affected by an inhibition of hard scattering.
Particles with mass larger than 246 GeV/c${}^2$ should only be observable through their 
gravitational effects, either contributing significantly to dark matter, or via 
detection of gravitational waves emitted during their collisions, provided that 
the propagation of gravitons is not subjected to the limitations of the horizon. 
On a final note, we observe that by attributing a finite size to elementary 
particles makes possible to discuss them in terms of energy density rather than 
energy in absolute terms. This makes even more natural the need for a consistent 
merging, via the quantum, between elementary particle physics and genuine general 
relativity, the latter being quintessentially related to the interplay between 
high-energy density phenomena and spacetime.

\end{document}